\documentclass[floats,eqsecnum,aps,epsf,graphicx,twocolumn
]{revtex4}

\usepackage{epsfig}

\catcode`\@=11

\begin{document}
\title{The liquid-crystalline smectic blue phases
}
\author{Eric Grelet}
\affiliation{Centre de Recherche Paul Pascal, CNRS UPR 8641,
Avenue Albert Schweitzer, F-33600 Pessac, France.}
\date{\today}

\begin{abstract}
Smectic blue phases (BP$_{Sm}$) are new mesophases of thermotropic
liquid crystals, which exhibit a {\it{double geometrical
frustration}}: the extension of chirality in three spatial
dimensions like the classical blue phases, and helical twist
competing with smectic order, as in the TGB phases. The existence
of a quasi-long range smectic order in BP$_{Sm}$ phases breaks the
cubic symmetry of classical blue phases. The symmetries of these
new phases have been determined by X-ray scattering and optical
polarizing microscopy experiments.
\end{abstract}

\maketitle

\section{\label{intro}Introduction}

\begin{figure}[tbh]
\centerline{\epsfig{file=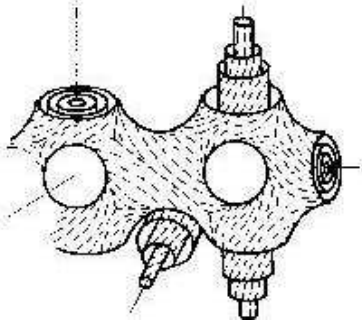,width=50mm}}
\caption{Geometrical model of blue phases involving double twist
cylinders and a network of disclination lines \cite{PRE}. }
\label{figBP}
\end{figure}

The discovery of new structures in liquid crystals turns out to be
often associated with an increase of complexity.  Smectic blue
phases are the most recent example combining the properties of two
types of frustrated mesophases: twist grain boundary (TGB) phases
and blue phases (BP). Blue phases, which were the first reported
liquid crystals \cite{reinitzer}, show an unusual cubic symmetry
in which the orientational order is periodic and long range in
three dimensions \cite{BP}. Thus blue phases are exemplary {\it
liquid crystals}: BPs exhibit three-dimensional cubic faceted
monodomains as in a crystal \cite{BP2}, but the order at the
origin of the cubic symmetry is not a positional order, but an
orientational one. 
The naming of these phases as {\it blue} is due to their Bragg
reflections in the visible range (and therefore also in the blue
wavelengths) indicating a spatially periodic structure with
lattice parameters of several hundreds nm. Saupe was the first to
suggest a chiral cubic structure from the BPs' optical activity
and their lack of birefringence ($\Delta$n=0) \cite{saupe}. The
BP's structure involves a radial twist of the director called a
double cylinder. However this double twisted structure cannot
extend perfectly into three-dimensional space. Geometrical models
of blue phases therefore consist of cubic networks of double twist
cylinders separated by disclination lines (Fig. \ref{figBP}).

\begin{figure}[tbh]
\centerline{\epsfig{file=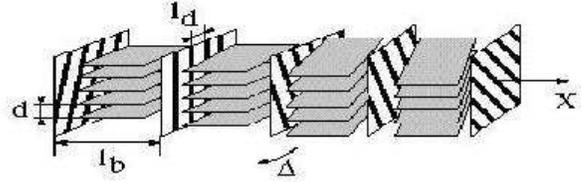,width=78mm}} \caption{Sketch
of the helical structure of the TGB$_{A}$ phase \cite{renn}. }
\label{figTGB}
\end{figure}

Twist grain boundary (TGB) phases represent a second example of a
frustrated chiral system. TGB phases were predicted by Renn and
Lubensky \cite{renn} and were experimentally observed by Goodby
{\it{et al.}} in 1989 for TGB$_{A}$ \cite{goodby} and by Nguyen
{\it{et al.}} in 1992 for TGB$_{C}$ \cite{thin1}. Since smectic
layers cannot be continuously twisted, the TGB phases consist of
blocks of pure smectic material (which can be either smectic-A for
TGB$_{A}$ or smectic-C for TGB$_{C}$) separated by parallel,
regularly spaced grain boundaries, formed by a periodic array of
screw dislocations (Fig. \ref{figTGB}). Such a dislocation
arrangement allows helical twist.

Thus, in TGB phases, as in blue phases, the frustration is
relieved by the presence of defects.\\

\begin{figure}[tbh]
\centerline{\epsfig{file=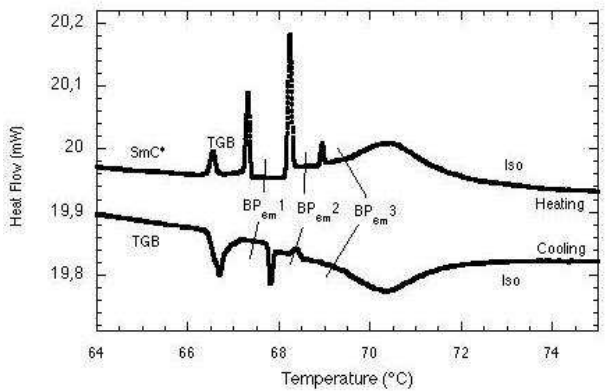,width=80mm}} \caption{ Typical
differential scanning calorimetry thermograms in the temperature
range of smectic blue phases performed by heating and by cooling
at 0.2$^{{\rm o}}$C/min \cite{LiqCrystEric}. } \label{DSC}
\end{figure}

In the last few years the Bordeaux group has synthesized new
chiral liquid crystals exhibiting mesophases called smectic blue
phases (BP$_{Sm}$), which appear in a very narrow temperature
range (Fig. \ref{DSC}) with the following phase sequence: TGB -
BP$_{Sm}$1 - BP$_{Sm}$2 - BP$_{Sm}$3 - Iso, without any
intermediate cholesteric state between the BP$_{Sm}$ and TGB
phases \cite{li2,LiqCrystEric,LiqCrystCristina}. The evidence of a
new phase sequence has stimulated   further experimental studies
by the Orsay group mainly based on optical techniques (polarizing
microscopy and optical activity measurements) and on X-ray
scattering from BP$_{Sm}$ monodomains.

\section{\label{Exp}The discovery of new mesophases}
\subsection{\label{optics}Optical experiments}

\begin{figure}[tbh]
\centerline{\epsfig{file=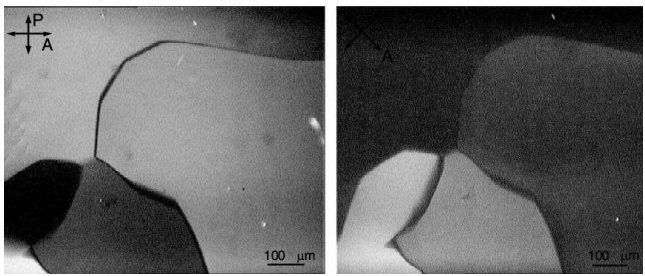,width=85mm}}
\caption{Birefringence of the BP$_{Sm}$2 phase observed in
transmission by polarizing microscopy. The thickness of the sample
is 100 microns. } \label{birefringenceBPSm2}
\end{figure}

\begin{figure}[tbh]
\centerline{\epsfig{file=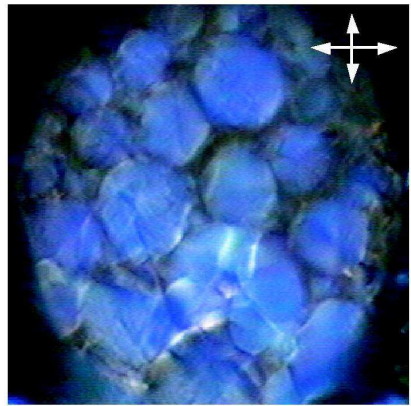,width=55mm}} \caption{
BP$_{Sm}$1 phase observed in transmission by polarizing
microscopy. BP$_{Sm}$1 phase is optically isotropic, that shows a
cubic symmetry, and the blue color is due to its optical activity.
The sample is 100 microns thick. } \label{textureBPSmA1}
\end{figure}

One of the first experiments to characterize mesophases is to
observe their texture under the polarizing microscope. The results
of such an experiment are reported in Figs.
\ref{birefringenceBPSm2} and \ref{textureBPSmA1} providing some
important information on the BP$_{Sm}$ structure.

When cooling from the isotropic phase, two smectic blue phases
with detectable textures occur in a narrow temperature span (about
1 $^{{\rm o}}$C): BP$_{Sm}$2 and BP$_{Sm}$1. The third smectic
blue phase (BP$_{Sm}$3) has, like the classical BP3 phase
\cite{coll}, an amorphous structure of the same macroscopic
symmetry as the isotropic phase \cite{ASCpeter}. As depicted in
Fig. \ref{birefringenceBPSm2}, a mosaic
texture composed of gray 
platelets is observed for BP$_{Sm}$2. This texture looks on first
view very similar to the one of classical blue phases~; however
what is the origin of the lack of bright colors observed in
smectic blue phases ? In BP's, the mosaic texture comes from
selective Bragg reflections at visible wavelengths from domains
with different orientations. In BP$_{Sm}$2 the existence of
modulation in colors from light gray to black is due to a
completely different optical feature: the birefringence
\cite{PRLEric}. BP$_{Sm}$2 phase is thus optically anisotropic and
the evidence of a non-cubic symmetry proves that smectic blue
phases are really new mesophases, not merely ``atypical'' blue
phases.

\begin{figure}[tbh]
\centerline{\epsfig{file=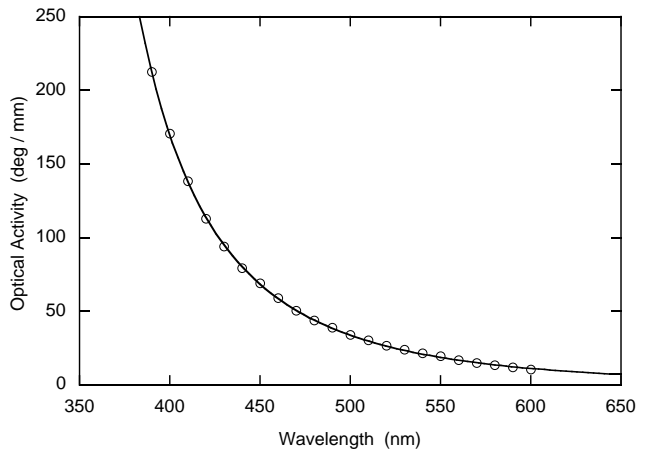,width=85mm}}
\caption{Optical activity as a function of wavelength for
BP$_{SmA}$1 (14FBTMHC ; T=72.8$^{{\rm o}}$C). The solid line is a
fit to
$\rho=(Cst/\lambda)/[\lambda^{2}({\lambda_{0}}^{2}-\lambda^{2})]$
where $\rho$ is the optical activity, $\lambda$=2$\pi$/k is the
wavelength of the light (k being the wavevector of the light), and
$\lambda$$_{0}$=nP (P being the lattice parameter and n the
average refractive index) \cite{articlePeter}. } \label{c14lambda}
\end{figure}

Contrary to the gray color of the BP$_{Sm}$2 phase, the blue color
of the BP$_{Sm}$1 phase shown in Fig. \ref{textureBPSmA1}
originates from its optical activity; this can only be seen in
absence of high birefringence \cite{MCLC}. Moreover, in smectic
blue phases the lattice parameter is in the near UV-range, and is
therefore too small to generate selective reflections of visible
light as in the BP's \cite{articlePeter}. The BP$_{Sm}$ lattice
parameter has been measured by studying the wavelength dependence
of the optical activity in BP$_{Sm}$1. A divergence of the optical
activity is expected close to the selective reflection,
$\lambda$$_{0}$, as has already been seen in the cholesteric phase
and in the classical blue phases. The data for the ``14BTMHC''
compound are plotted in Fig. \ref{c14lambda}, showing a rise of
the optical activity when the wavelength decreases. A fit allows
an estimate of the BP$_{Sm}$ lattice parameter: P=0.21 ${\mu}m$
with $\lambda$$_{0}$=320 nm and n=1.5.

The value of the BP$_{Sm}$ lattice parameter in the UV range
prevents study by optical scattering of visible light (Kossel
diagram technique), which has been used to determine the symmetry
of classical blue phases \cite {kossel}. Therefore, other ways had
to be found to study the orientational symmetries of smectic blue
phases, such as the growth of faceted crystallites.

\begin{figure}[tbh]
\centerline{\epsfig{file=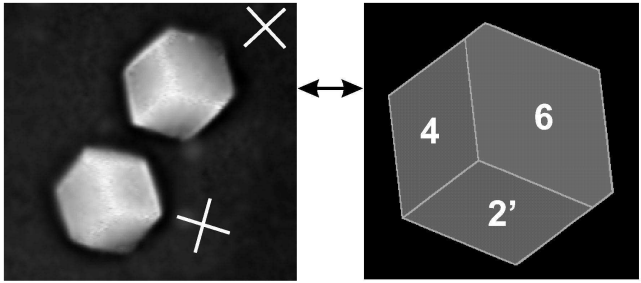,width=85mm}} \caption{Experimental
and schematic view of BP$_{Sm}$2 monocrystals having almost the
same orientation floating in the isotropic supercooled BP$_{Sm}$3
and observed along a pseudothreefold axis in transmission between
crossed polarizers \cite{PREfacet}. The white cross represents the
projection (or the normal direction) of the optical axis in the
observation plane.} \label{A3}
\end{figure}

\begin{figure}[tbh]
\centerline{\epsfig{file=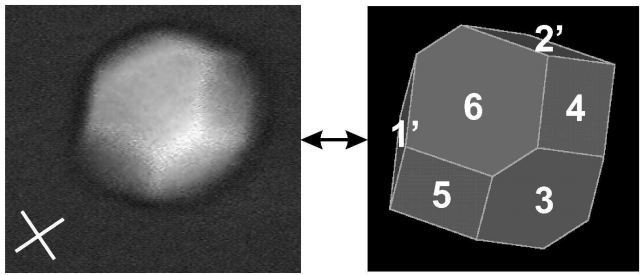,width=85mm}} \caption{Faceted
monodomain of the BP$_{Sm}$2 phase exhibiting four large and two
small facets \cite{PREfacet}. } \label{A4}
\end{figure}

\begin{figure}[tbh]
\centerline{\epsfig{file=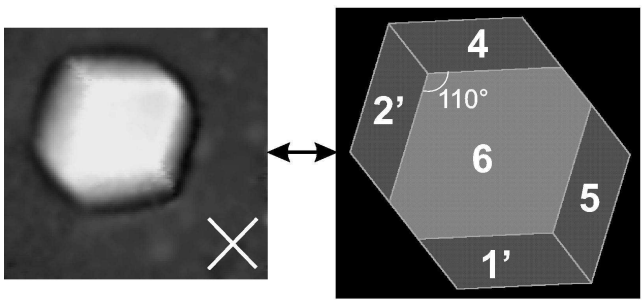,width=85mm}} \vspace{2mm}
\caption{BP$_{Sm}$2 monocrystal floating in the supercooled
BP$_{Sm}$3 and observed along a two-fold axis in transmission
between crossed polarizers \cite{PREfacet}.} \label{A2}
\end{figure}

\begin{figure}[tbh]
\centerline{\epsfig{file=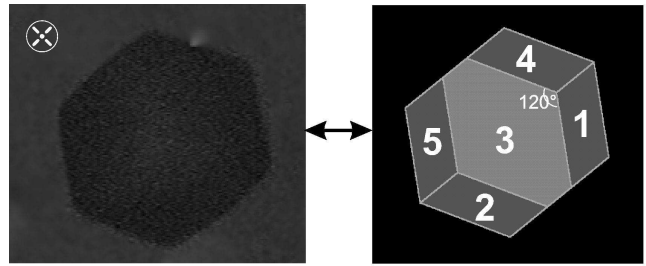,width=85mm}} \vspace{2mm}
\caption{Crystal shape of BP$_{Sm}$2 which stays dark for all the
positions of the crossed polarizers \cite{PREfacet}. The optical
axis is then perpendicular to the plane of the figure. An angle of
about 120${{}^{\circ }}$ is observed on this picture. } \label{A6}
\end{figure}

The nucleation and growth of such single faceted crystals of
BP$_{Sm}$ phases represents a real experimental challenge. Using a
low cooling rate (0.01 $^{{\rm o}}$C per 10 min) to produce large
monodomains gives rise to the birefringent platelet texture (Fig.
\ref{birefringenceBPSm2}) that quickly fills the entire
experimental cell. However, by using the metastability of the
BP$_{Sm}$3 phase occurring in some compounds, we succeeded in
producing large faceted crystallites of the BP$_{Sm}$2 phase
\cite{PREfacet}. This kind of experiment is difficult because it
is performed out of equilibrium in a very narrow temperature range
(0.15 $^{{\rm o}}$C). Different crystalline shapes have been
listed, as reported in Figs. \ref{A3} to \ref{A6}. These large
monocrystals, between 100 and 150 ${\mu}m$ in size and with
well-defined facets, have been observed floating in the bulk in
coexistence with the isotropic supercooled BP$_{Sm}$3 phase. On
each picture, the white cross represents the positions of the
crossed polarizers needed to assure {\it{extinction}} of the
birefringent domains, i.e., it represents the projection (or the
normal direction) of the optical axis in the observation plane.
The three-dimensional polyhedral habit of the BP$_{Sm}$2
monocrystals shown in Figs. \ref{A3} to \ref{A6} seems close to a
rhombic dodecahedron, which has already been observed for
classical cubic blue phase \cite{BP2}. Nevertheless, the
experimental crystallites cannot reproduce a perfect rhombic
dodecahedron because their birefringence is incompatible with a
cubic structure.

\begin{figure}[tbh]
\centerline{\epsfig{file=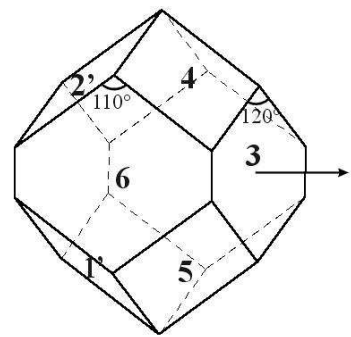,width=45mm}}
\caption{Orthorhombic dodecahedral crystal habit formed by four
(100) facets (labeled 3 and 6) and eight (111) facets (labeled 1,
2, 4, and 5) accounting for the experimentally observed
monocrystals of the BP$_{Sm}$2 phase \cite{PREfacet}. For this
model, the parameters of the unit cell are a=1, b=0.82, and c=0.58
and the optical axis is located perpendicular to the normal of the
facet labeled 3. } \label{orthorhombic}
\end{figure}

Thus, the data of the birefringent faceted monodomains suggest
that the crystal habit of the BP$_{Sm}$2 phase is formed by a
dodecahedral structure made from an orthorhombic unit cell (Fig.
\ref{orthorhombic}). This interpretation of an orthorhombic
symmetry is also consistent with the X-ray scattering experiments
and especially with the epitaxial relation observed at the
transition from BP$_{Sm}$2 to BP$_{Sm}$1 phases
\cite{PREfacet,MCLC}.

\subsection{\label{RX}X-ray scattering experiments}

\begin{figure}[tbh]
\centerline{\epsfig{file=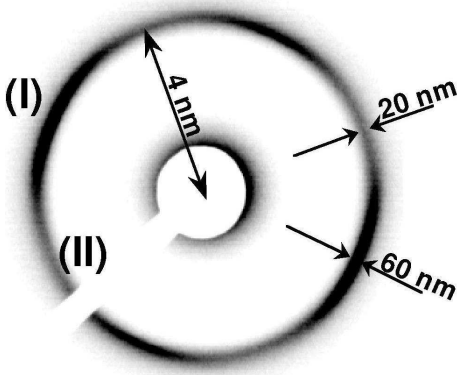,width=60mm}}
\caption{Experimental x-ray scattering patterns of BP$_{Sm}$2
monodomains. Parts of the ring with higher intensity (labeled
(I)), where the smectic order is enhanced, correspond to the
``Bragg peaks'', and parts with lower intensity are marked (II).
The layer spacing and the correlation lengths of the smectic order
are also indicated \cite{PRLEric,PREBPSmC1}.} \label{RX-pattern}
\end{figure}

The new phase sequence (TGB - BP$_{Sm}$ - Iso) has stimulated
X-ray scattering studies, especially to investigate whether the
smectic order existing in the TGB phases persists in smectic blue
phases. The experiments show that the BP$_{Sm}$ phases exhibit,
contrary to BPs, quasi-long-range smectic order with a typical
persistence length of about 60 nm. Information on the symmetry of
these new phases has then been obtained in growing BP$_{Sm}$
monodomains to explore the entire reciprocal space
\cite{PRE,PRLEric,PREBPSmC1}. The scattering patterns obtained
exhibit pairs of ``Bragg peaks'' indicating that the smectic order
is not isotropic, but extends in given directions of the
three-dimensional unit cell (Fig. \ref{RX-pattern}). By combining
these various profiles, the directions of smectic order
enhancement i.e. those in which the different ``Bragg peaks'' are
located, have been determined and  the angles between these
directions have been deduced, providing the symmetry of each
BP$_{Sm}$ phase.

It is important to note here that the Bragg scattering is not due
to the periodicity of the orientational order, which is at the
length scale of the unit cell (about 200 nm, the dimension of the
lattice parameter) (Fig. \ref{c14lambda}) \cite{articlePeter}.
Only the smectic order has been probed, defined by a periodicity
of 4 nm and a correlation length of about 60 nm, three times
smaller than the size of the unit cell (Fig. \ref{RX-pattern}).
Therefore, these x-ray scattering experiments only provide {\it
indirect} information on the symmetry of the orientational unit
cell, contrary to the faceted monocrystals of the BP$_{Sm}$2
phase.

These results by X-ray scattering point to the main unsolved
theoretical question dealing with smectic blue phases: why does
the smectic order break the cubic symmetry of blue phases ?

\section{\label{models} Geometrical models of smectic blue phases}

\begin{figure}[tbh]
\centerline{\epsfig{file=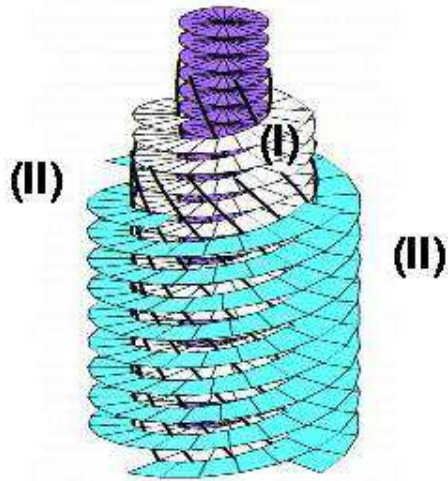,width=60mm}}
\caption{Geometrical model of a smectic double twist cylinder
giving rise to the experimental smectic peaks (called region (I)
in Fig. \ref{RX-pattern}). In this picture, the surfaces represent
the smectic layers and the black lines wrapping around the
cylinder represent the screw dislocations. However, smectic order
persists between cylinders and gives rise to the continuous
smectic ring (region labeled (II) in Fig. \ref{RX-pattern})
\cite{kamien}. } \label{KamienModel}
\end{figure}

The first theoretical approach for combining smectic order with
three-dimensional orientational order was proposed by Kamien
\cite{kamien} with a model of smectic double twist cylinders
linking smectic A order and twist not only in one direction as in
TGB phases, but in two spatial directions (Fig.
\ref{KamienModel}). The smectic double twist cylinder is the
counterpart in the smectic blue phases of the double twist
cylinder in the BP's.

Our experimental results can be interpreted in terms of this
geometrical model by assuming that the regions where the smectic
order easily extends, are the smectic double twist cylinders cores
and correspond to the ``Bragg peaks'' in Fig. \ref{RX-pattern}. A
geometrical model of the structure of smectic blue phases can be
sketched by packing these smectic double twist cylinders according
to the observed symmetries \cite{PRE}\cite{PRLEric}. Thus some
topological defects exist both {\it in} the smectic double twist
cylinder for combining twist and smectic order, and {\it between}
the cylinders as in blue phases.

Recently another geometrical model based on minimal surfaces has
also been suggested \cite{kamien-new}.

\section{\label{Conclusion}Conclusion}

In the same way as a classical blue phase can be seen as the
three-dimensional counterpart of the cholesteric phase, a smectic
blue phase can be regarded as the extension to 3D of the TGB
phase. Indeed, contrary to the cholesteric and to the blue phases
where the twist occurs at the ``molecular level'', the twist
occurs at the scale of the smectic slabs for both TGB and smectic
blue phases. Thus smectic blue phases represent an original
physical system of condensed matter exhibiting a {\it{double
frustration}}: the extension of chirality in three spatial
dimensions, and helical twist competing with smectic order.

\begin{acknowledgments}
Peter J. Collings, Randall D. Kamien, Min-Hui Li, Huu Tinh Nguyen
and Brigitte Pansu are acknowledged for their contributions to the
research on the smectic blue phases.
\end{acknowledgments}

\vfill \clearpage

\end{document}